\documentstyle[prd,tighten,aps,floats,amssymb,newlfont,graphicx]{revtex}
\begin{document}
\draft
\twocolumn[\hsize\textwidth\columnwidth\hsize\csname
@twocolumnfalse\endcsname
\title{Four-neutrino oscillations and the solar neutrino problem}
\author{C. Giunti$^{1}$, M.C. Gonzalez-Garcia$^{2}$ and 
C. Pe\~na-Garay$^{2}$} 
\address{\sl
$^1$ INFN, Sez. di Torino, and Dip. di Fisica Teorica, 
Univ. di Torino, I-10125 Torino, Italy\\ 
$^2$  Instituto de F\'{\i}sica Corpuscular (IFIC)
CSIC-Universidad de Valencia,\\
Edificio Institutos de Paterna, Apartado 2085, 46071 Valencia}
\date{July 14, 2000}
\maketitle
\begin{abstract}
We perform a fit
of solar neutrino data
in the framework of
the two four-neutrino schemes
that are compatible with the results of all neutrino oscillation experiments.
These schemes
allow simultaneous transitions
of solar $\nu_e$'s into active $\nu_\mu$'s, $\nu_\tau$'s
and sterile $\nu_s$.
The data imply that the SMA solution
is valid for any combination of
$\nu_e$$\to$active
and
$\nu_e$$\to$sterile transitions,
whereas the LMA, LOW and VO
solutions disappear when
$\nu_e\to\nu_s$
transitions are dominant.
\\
\sf
DFTT 29/00, arXiv:hep-ph/0007154.
\\
\sf
Talk presented by C. Giunti at the
NuFact'00 Workshop,
May 22-26, 2000,
Naval Postgraduate School,
Monterey, California, USA.
\end{abstract}
\pacs{}
]

\section{Introduction:\\why four-neutrino mixing?}
\label{intro}

Evidences in favor of neutrino oscillations
(see \cite{BGG-review-98})
have been
found in solar
(Homestake \cite{Homestake-98},
Kamiokande \cite{Kamiokande-sun-96},
GALLEX \cite{GALLEX-99},
SAGE \cite{SAGE-99},
Super-Kamiokande
\cite{SK-sun-lp99})
and atmospheric
(Kamiokande, IMB, Super-Kamiokande, Soudan 2, MACRO
\cite{exp-atm})
neutrino experiments
and in the LSND experiment
\cite{LSND}.
It is rather well-known
\cite{Giunti-JHU-99,Fogli-Lisi-Marrone-Scioscia-no3-99}
that
these three evidences
require the existence of
at least three small neutrino mass-squared differences
($\Delta{m}^2$'s):
\begin{eqnarray}
&
\Delta{m}^2_{\mathrm{sun}} \sim 10^{-10} - 10^{-4} \, \mathrm{eV}^2
\,,
&
\label{dm2-sun}
\\
&
\Delta{m}^2_{\mathrm{atm}} \sim 10^{-3} - 10^{-2} \, \mathrm{eV}^2
\,,
&
\label{dm2-atm}
\\
&
\Delta{m}^2_{\mathrm{LSND}} \sim 10^{-1} - 10 \, \mathrm{eV}^2
\,.
&
\label{dm2-LSND}
\end{eqnarray}
This means that at least four light massive neutrinos must exist in nature.
Following the old principle known as
Occam's razor
(``pluralitas non est ponenda sine necessitate'')
we assume the existence of four light massive neutrinos,
that is the minimal possibility compatible with all data.
In this case 
the left-handed components
$\nu_{{\alpha}L}$
of the flavor neutrino fields
are superpositions of
the left-handed components
$\nu_{kL}$
($k=1,\ldots,4$)
of neutrino fields with definite mass
$m_k$ smaller than a few eV:
\begin{equation}
\nu_{{\alpha}L}
=
\sum_{k=1}^{4}
U_{{\alpha}k}
\,
\nu_{kL}
\,,
\label{mixing}
\end{equation}
where $U$
is a $4{\times}4$ unitary mixing matrix.

From the measurement of the invisible decay width of the $Z$-boson
it is known that the number of light active
neutrino flavors is three
(see \cite{PDG-98}),
corresponding to $\nu_e$, $\nu_\mu$ and $\nu_\tau$
(active neutrinos are
those taking part to standard weak interactions).
This implies that
the number of massive neutrinos is bigger or equal to three.
If there are four massive neutrinos,
in the flavor basis there is one sterile neutrino,
$\nu_{s}$,
that does not take part to standard weak interactions.
For the flavor index $\alpha$ in Eq.~(\ref{mixing})
we choose the ordering
$\alpha=e,s,\mu,\tau$.

Let us emphasize that four-neutrino mixing
is very interesting both for theory and experiment.
From the theoretical point of view
the existence of a light sterile neutrino
can be explained only with models
far beyond the Standard Model
and is thus a signal of exciting new physics.
From the experimental point of view
four-neutrino mixing
is attractive
because it allows the existence
of a $\Delta{m}^2$ at the eV scale
which generates oscillations
in different channels that can be explored with high
precision in short-baseline experiments.
Moreover,
it allows large CP-violation effects
in long-baseline and $\nu$-factory experiments
(see \cite{Donini-nufact00}).
Four-neutrino mixing may also open the possibility
for measurable exotic phenomena,
as neutrino decay, neutrino magnetic moment, etc.

From the model-builder point of view,
there seems to be no difficulty
in constructing models with a light sterile neutrino
(as one can convince oneself with an appropriate search of
the hep-ph electronic archive,
resulting in too many four-neutrino models to be cited here).
However,
our impression is that
these models are constructed with assumptions ``ad hoc''
in order to generate a light sterile neutrino,
with the possible exception of
models based on the existence of a mirror world
(see \cite{mirror-world})
and models based on the existence of large extra dimensions
(see \cite{Neubert-nufact00}),
in which light neutral fermions that mix with
the ordinary neutrinos seem to be naturally allowed.

Our approach is phenomenological:
we extract information on four-neutrino mixing
(masses and mixing angles)
from the available data.

\section{Allowed four-neutrino schemes}
\label{allowed}

\begin{figure}[t]
\begin{center}
\includegraphics[bb=13 668 522 827,width=\linewidth]{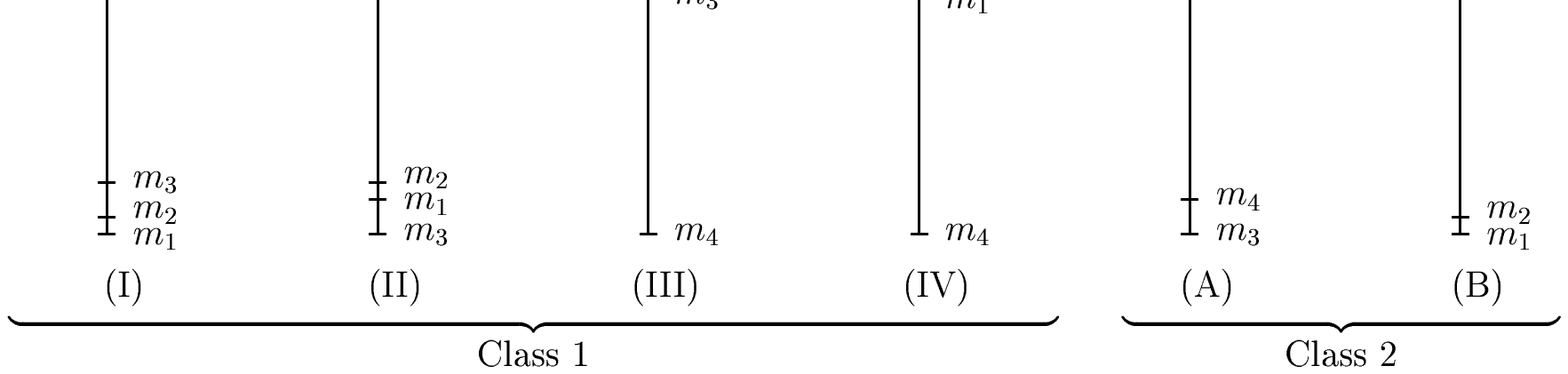}
\refstepcounter{figure}
\label{4schemes}
\\
\small
Figure \ref{4schemes}
\end{center}
\end{figure}

The six types of four-neutrino mass spectra
with
three different scales of $\Delta{m}^2$
that can accommodate
the hierarchy
$
\Delta{m}^2_{\mathrm{sun}}
\ll
\Delta{m}^2_{\mathrm{atm}}
\ll
\Delta{m}^2_{\mathrm{LSND}}
$
(see Eqs.~(\ref{dm2-sun})--(\ref{dm2-LSND}))
are shown qualitatively in Fig.~\ref{4schemes}.
In all these mass spectra there are two groups
of close masses separated by the ``LSND gap'' of the order of 1 eV.
The six schemes are divided into four schemes of class 1 (I--IV)
in which there is a group of three masses separated from an isolated mass
by the LSND gap,
and two schemes of class 2 (A, B)
in which there are two couples of close masses separated by the LSND gap.
In each scheme the smallest mass-squared difference
($\Delta{m}^2_{ij} \equiv m_i^2 - m_j^2$)
corresponds to
$\Delta{m}^2_{21} = \Delta{m}^2_{\mathrm{sun}}$,
the intermediate one to
$\Delta{m}^2_{\mathrm{atm}}$
($\Delta{m}^2_{31} \simeq \Delta{m}^2_{32}$
in the schemes of class 1,
$\Delta{m}^2_{43}$
in the schemes of class 2)
and the largest mass squared difference
$\Delta{m}^2_{41} = \Delta{m}^2_{\mathrm{LSND}}$
is relevant for the oscillations observed in the LSND experiment
($\Delta{m}^2_{41} \simeq \Delta{m}^2_{42} \simeq \Delta{m}^2_{43}$
in the schemes of class 1,
$\Delta{m}^2_{41} \simeq \Delta{m}^2_{42} \simeq
\Delta{m}^2_{31} \simeq \Delta{m}^2_{32}$
in the schemes of class 2).

It has been show that
the four schemes of class 1 are disfavored by the data
if also the negative
results of short-baseline accelerator and reactor disappearance
neutrino oscillation experiments are taken into account
\cite{BGG-AB-96,Barger-variations-98,BGGS-AB-99,Giunti-JHU-99}.
The reason is simple:
in order to explain solar and atmospheric data
with neutrino oscillations
$\nu_e$ and $\nu_\mu$
must have large mixing with the group of three neutrinos
and, consequently,
small mixing with the isolated neutrino.
In this case,
short-baseline oscillations
occurring through the mass-squared difference $\Delta{m}^2_{41}$,
\textit{i.e.}
through the interference of the wave functions
of the isolated neutrino and the three grouped neutrinos,
are strongly suppressed.
It turns out that
the negative
results of short-baseline accelerator and reactor disappearance
experiments imply that
the amplitude of short-baseline $\nu_\mu\to\nu_e$
transitions is suppressed below the value measured by LSND
\cite{BGG-AB-96,BGGS-AB-99,Giunti-JHU-99}.
Hence,
the schemes of class 1 cannot explain all neutrino oscillation data. 

\begin{figure}[t]
\begin{center}
\includegraphics[bb=98 644 513 773,width=\linewidth]{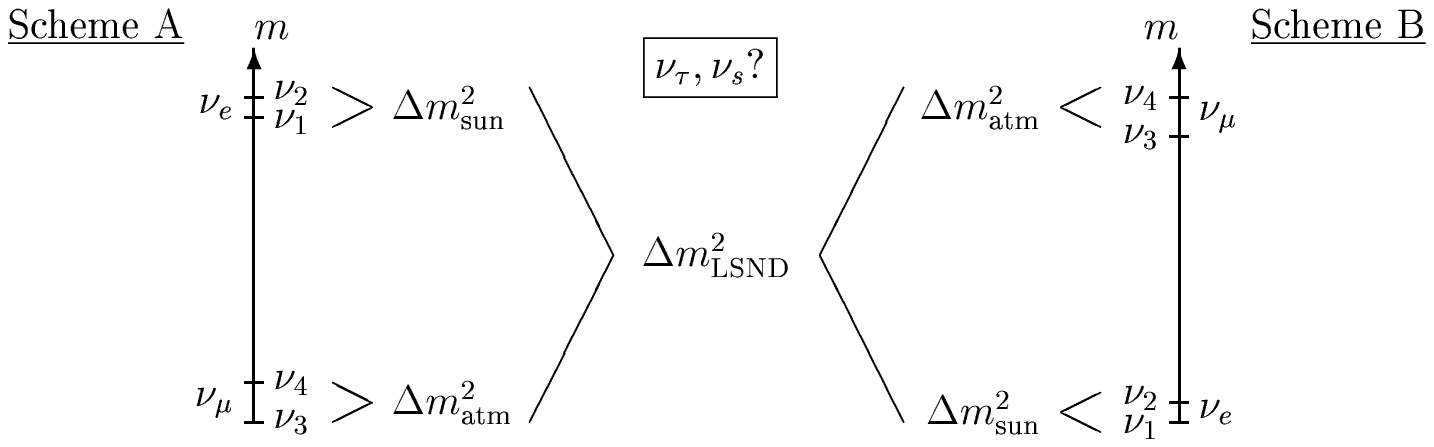}
\refstepcounter{figure}
\label{4schemesAB}
\\
\small
Figure \ref{4schemesAB}
\end{center}
\end{figure}

On the other hand,
the two four-neutrino schemes of class 2
are compatible with the results of all neutrino oscillation experiments
if
the mixing of $\nu_e$ with
$\nu_1$ and $\nu_2$,
the two mass eigenstates responsible
for the oscillations of solar neutrinos,
and
the mixing of $\nu_\mu$ with
$\nu_3$ and $\nu_4$,
the two mass eigenstates responsible
for the oscillations of atmospheric neutrinos,
are large
\cite{BGG-AB-96,Barger-variations-98,BGGS-AB-99,Giunti-JHU-99}.
Hence,
we have the two possibilities illustrated in Fig.~\ref{4schemesAB},
in which $\nu_e$ and $\nu_\mu$
are depicted close to the two massive neutrinos
with which they have large mixing
and it is explicitly indicated that
the mixing of $\nu_\tau$ and $\nu_s$ is unknown.
Information on the mixing of $\nu_\tau$ and $\nu_s$,
as well as more detailed information on the mixing of $\nu_e$ and $\nu_\mu$,
can be obtained through the fit of solar and atmospheric neutrino data,
and,
in the future,
of the data of long-baseline experiments.

The appropriate formalism for the treatment of
solar and atmospheric neutrino oscillations,
including matter effects,
in the framework of the two four-neutrino schemes A and B
has been presented in Ref.~\cite{DGKK-99}.
In the next section we present the results of a fit of solar neutrino data
\cite{Concha-foursolar-00}.

\section{Fit of solar neutrino data}
\label{fit}

\begin{figure}[t]
\begin{center}
\includegraphics[bb=125 700 483 759,width=\linewidth]{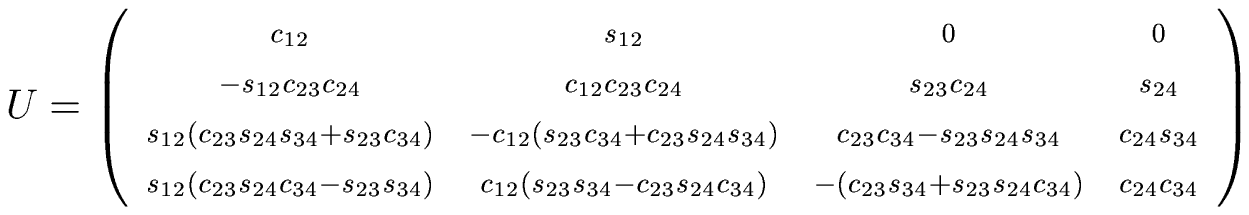}
\refstepcounter{figure}
\label{mixing_matrix}
\\
\small
Figure \ref{mixing_matrix}
\end{center}
\end{figure}

In the framework of the two four-neutrino schemes A and B,
solar neutrino oscillations depend on only one mass-squared difference,
$\Delta{m}^2_{\mathrm{sun}}$,
but there are several degrees of freedom coming from the
$4\times4$
mixing matrix,
that can be parameterized in terms of six mixing angles
and three CP-violating phases
(three additional CP-violating phases are possible if
massive neutrinos are Majorana particles,
but can be neglected in the present contest because they do not
have any effect in neutrino oscillations).
However,
the data of solar neutrino experiments
together with those of
short-baseline $\bar\nu_e$ disappearance experiments
indicate that the elements
$U_{e3}$ and $U_{e4}$
of the neutrino mixing matrix
are very small
\cite{BGG-AB-96,Giunti-JHEP-00}.
This implies that two mixing angles are small
and can be neglected in the study
of solar (and atmospheric) neutrino oscillations
\cite{DGKK-99}.
Furthermore,
the CP-violating phases can be neglected because their
effects are washed out by the average over neutrino energy
and source-detector distance.
Therefore,
we consider the mixing matrix $U$ in Fig.~\ref{mixing_matrix}
\cite{DGKK-99,Concha-foursolar-00},
where 
$\vartheta_{12}$, 
$\vartheta_{23}$, 
$\vartheta_{24}$, 
$\vartheta_{34}$ 
are four mixing angles 
and 
$ c_{ij} \equiv \cos\vartheta_{ij} $ 
and 
$ s_{ij} \equiv \sin\vartheta_{ij} $. 
The rows and columns of the mixing matrix $U$
correspond,
respectively,
to the neutrino flavors $\nu_{e},\nu_{s},\nu_{\mu},\nu_{\tau}$
and
to the mass eigenstates $\nu_{1},\nu_{2},\nu_{3},\nu_{4}$.

Since solar neutrino oscillations 
are generated by the mass-square difference 
between $\nu_2$ and $\nu_1$
and
$U_{e1} = \cos\vartheta_{12}$,
$U_{e2} = \sin\vartheta_{12}$,
the survival of solar $\nu_e$'s,
$P_{\nu_e\to\nu_e}$,
mainly depends on the mixing angle
$\vartheta_{12}$,
with the only correction due to four-neutrino mixing
in the matter potential that is given by
\begin{equation}
V
\equiv
V_{CC} + \cos^2{\vartheta_{23}} \, \cos^2{\vartheta_{24}} \, V_{NC}
\,,
\label{V}
\end{equation}
where
$V_{CC} = \sqrt{2} G_F N_e$
and
$V_{NC} = - \sqrt{2} G_F N_n / 2$
are the usual charged-current and neutral-current potentials
(see \cite{BGG-review-98}).

The mixing angles
$\vartheta_{23}$ and $\vartheta_{24}$ 
determine the relative amount of transitions into sterile $\nu_s$ 
or 
active $\nu_\mu$ and $\nu_\tau$
(that cannot be distinguished in solar neutrino experiments, 
because their matter potential 
and their interaction in the detectors are equal, 
due only to neutral-current weak interactions). 
The active/sterile ratio 
and solar neutrino oscillations in general 
do not depend on 
the mixing angle 
$\vartheta_{34}$, 
that contribute only to the different mixings of 
$\nu_\mu$ and $\nu_\tau$.
Indeed, 
the mixing of 
$\nu_s$ with $\nu_1$ and $\nu_2$ 
depends only on 
$\vartheta_{12}$ 
and the product 
$\cos{\vartheta_{23}} \cos{\vartheta_{24}}$. 
Moreover, 
instead of 
$\nu_\mu$ and $\nu_\tau$, 
one can consider 
the linear combinations 
$$
\left( 
\begin{array}{l} 
\nu_a 
\\ 
\nu_b 
\end{array} 
\right) 
= 
\left( 
\begin{array}{cc} 
- \sin{\vartheta} & - \cos{\vartheta} 
\\ 
\cos{\vartheta} & - \sin{\vartheta} 
\end{array} 
\right) 
\left( 
\begin{array}{cc} 
s_{34} & c_{34} 
\\ 
c_{34} & - s_{34} 
\end{array} 
\right) 
\left( 
\begin{array}{l} 
\nu_\mu 
\\ 
\nu_\tau 
\end{array} 
\right) 
\,, 
$$
with 
$
\tan\vartheta 
= 
\sin{\vartheta_{24}} / \tan{\vartheta_{23}}
$.
The mixing of 
$\nu_a$ and $\nu_b$ 
with 
$\nu_1$ and $\nu_2$ 
is given by
$
U_{a1} 
= 
- s_{12} 
\sqrt{1 - c_{23}^2 c_{24}^2} 
$,
$
U_{a2} 
= 
c_{12} 
\sqrt{1 - c_{23}^2 c_{24}^2} 
$,
$
U_{b1} = U_{b2} = 0 
$.
Therefore, the oscillations of solar neutrinos depend only on 
$\vartheta_{12}$ and the product 
$\cos{\vartheta_{23}} \cos{\vartheta_{24}}$. 
If $0 < \cos{\vartheta_{23}} \cos{\vartheta_{24}} < 1$, 
solar $\nu_e$'s can transform
simultaneously
into sterile neutrinos
and
in the linear combination $\nu_a$ of active $\nu_\mu$ and $\nu_\tau$,
with the limiting cases of
pure two-generation 
$\nu_e\to\nu_a$ transitions
if $\cos{\vartheta_{23}} \cos{\vartheta_{24}} = 0$ 
($U_{s1}=U_{s2}=0$, 
$ 
U_{a1} 
= 
- \sin{\vartheta_{12}} 
$, 
$ 
U_{a2} 
= 
\cos{\vartheta_{12}} 
$), 
and
pure two-generation 
$\nu_e\to\nu_s$ transitions
if $\cos{\vartheta_{23}} \cos{\vartheta_{24}} = 1$
($ 
U_{s1} 
= 
- \sin{\vartheta_{12}} 
$, 
$ 
U_{s2} 
= 
\cos{\vartheta_{12}} 
$ 
and 
$U_{a1}=U_{a2}=0$).
 
Since 
the mixing of $\nu_e$ with $\nu_1$ and $\nu_2$ 
is equal to the one in the case of two-generations 
(with the mixing angle $\vartheta_{12}$), 
the mixing of $\nu_s$ with $\nu_1$ and $\nu_2$ 
is equal to the one in the case of two-generations 
times $\cos{\vartheta_{23}} \cos{\vartheta_{24}}$ 
and 
the mixing of $\nu_a$ with $\nu_1$ and $\nu_2$ 
is equal to the one in the case of two-generations 
times 
$\sqrt{ 1 - \cos^2{\vartheta_{23}} \cos^2{\vartheta_{24}} }$, 
it is clear that 
in the general case of simultaneous 
$\nu_e\to\nu_s$ 
and 
$\nu_e\to\nu_a$ 
oscillations 
the corresponding transition probabilities are given by 
\begin{eqnarray} 
&& 
P_{\nu_e\to\nu_s} 
= 
\cos^2{\vartheta_{23}} \cos^2{\vartheta_{24}} 
\left( 1 - P_{\nu_e\to\nu_e} \right) 
\,, 
\label{Pes} 
\label{Pea} 
\\ 
&& 
P_{\nu_e\to\nu_a} 
= 
\left( 1 - \cos^2{\vartheta_{23}} \cos^2{\vartheta_{24}} \right) 
\left( 1 - P_{\nu_e\to\nu_e} \right) 
\,. 
\end{eqnarray} 
These expressions satisfy the relation of 
probability conservation 
$ 
P_{\nu_e\to\nu_e} 
+ 
P_{\nu_e\to\nu_s} 
+ 
P_{\nu_e\to\nu_a} 
= 
1 
$.
We calculated the survival probability
$P_{\nu_e\to\nu_e}$
in the range
$
10^{-11} \, \mathrm{eV}^2
\leq
\Delta{m}^2
\leq
10^{-13} \, \mathrm{eV}^2
$,
$10^{-4} \leq \tan^2 \vartheta_{12} \leq 10$
and
$0 \leq \cos^2{\vartheta_{23}} \cos^2{\vartheta_{24}} \leq 1$,
taking into account
matter effects for
$
10^{-8} \, \mathrm{eV}^2
\leq
\Delta{m}^2_{21}
\leq
10^{-3} \, \mathrm{eV}^2
$
and
the possible regeneration of $\nu_e$'s
when the flux of solar neutrinos crosses the Earth
(for details see Ref.~\cite{Concha-foursolar-00}).

In order to
calculate the allowed regions for the parameters
$\Delta{m}^2_{21}$,
$\tan^2\vartheta_{12}$
and
$c^2_{23} c^2_{24}$,
we have used data 
on the total event rates measured in the Chlorine experiment at 
Homestake \cite{Homestake-98},
in the two Gallium experiments GALLEX \cite{GALLEX-99} and 
SAGE \cite{SAGE-99}
and in the water Cherenkov detectors
Kamiokande \cite{Kamiokande-sun-96}
and 
Super-Kamiokande
\cite{SK-sun-lp99}
shown in Table~\ref{rates}.
We used also
the zenith angle distribution of the events and the electron recoil 
energy spectrum measured in the Super-Kamiokande experiment
(825-day data sample)
\cite{SK-sun-lp99}.
For the  calculation of the theoretical expectations we use the BP98 standard  
solar model of Ref.~\cite{BP98}.
The calculation of the theoretical rates,
the statistical treatment of the data
and
the results of different fits of rates only,
rates and zenith angle distribution,
rates and recoil electron energy spectrum
are described in Ref.~\cite{Concha-foursolar-00}.
Here we present only the result of the global $\chi^2$ fit
of all data.

\begin{table}[t]
\begin{center}
\begin{tabular}{|l|c|c|c|} 
Experiment & Rate & Units& BP98 \\ 
\hline 
Homestake & $2.56\pm 0.23 $ \protect\cite{Homestake-98} & SNU &  $7.8\pm 1.1 $   \\ 
GALLEX+SAGE & $72.3\pm 5.6 $ \protect\cite{GALLEX-99,SAGE-99} & SNU & $130\pm 7 $  \\ 
Kamiokande & $2.80\pm 0.38$ \protect\cite{Kamiokande-sun-96} &  
$10^{6}$\,cm$^{-2}$s$^{-1}$ & $5.2\pm 0.9 $ \\    
Super-Kamiokande & $2.45\pm 0.08$ \protect\cite{SK-sun-lp99} &  
$10^{6}$\,cm$^{-2}$s$^{-1}$ & $5.2\pm 0.9 $ \\    
\end{tabular}
\refstepcounter{table}
\label{rates}
\small
Table \ref{rates}
\end{center}
\end{table} 

\begin{figure}[t]
\begin{center}
\includegraphics[bb=38 43 510 766,width=\linewidth]{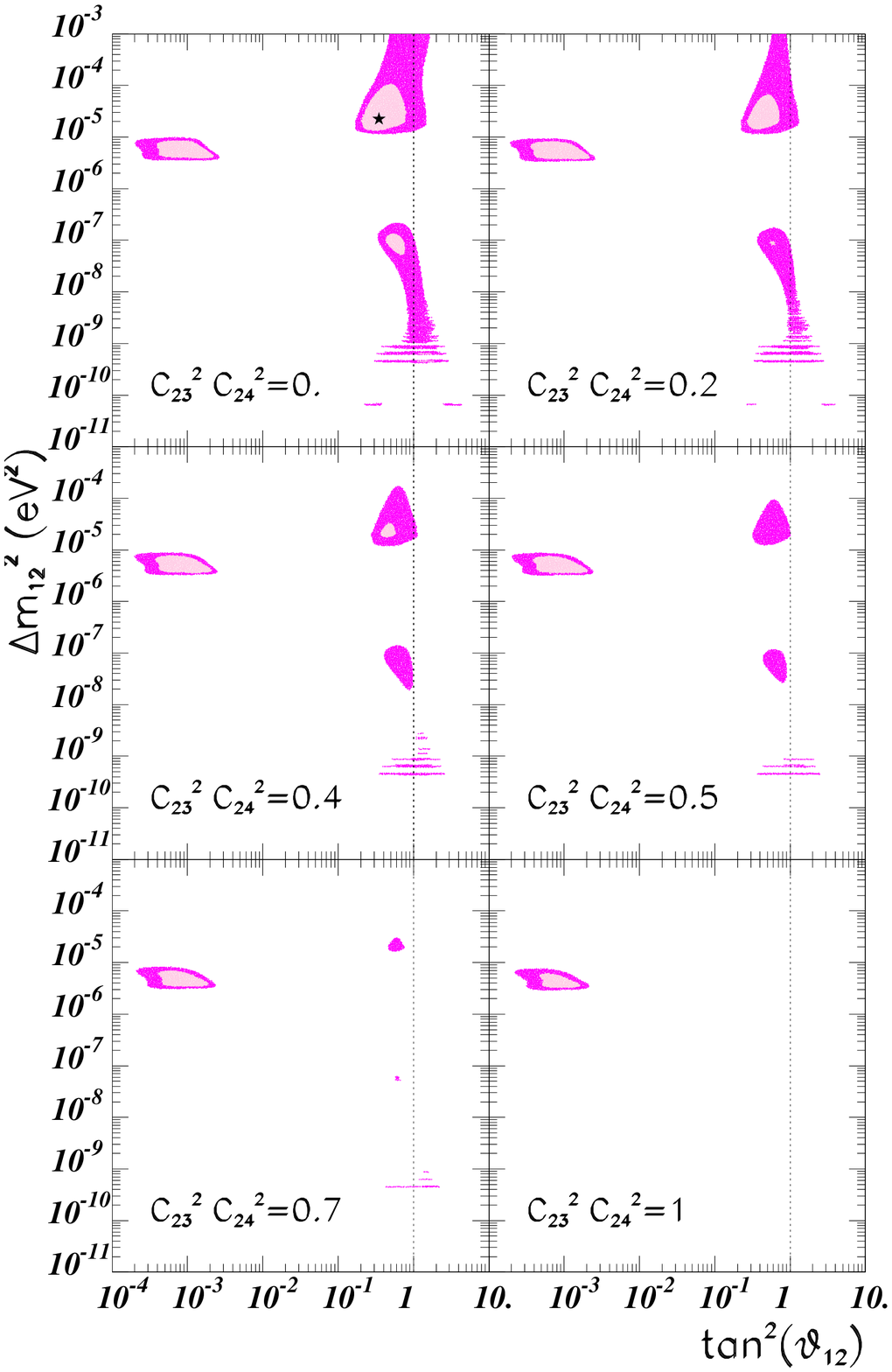}
\refstepcounter{figure}
\label{four}
\\
\small
Figure \ref{four}
\end{center}
\end{figure}

Figure~\ref{four}
shows the allowed regions at 90\% and 99\% CL in the
$\tan^2\vartheta_{12}$--$\Delta{m}^2_{21}$
plane for
$c^2_{23} c^2_{24} = 0, 0.2, 0.4, 0.5, 0.7 1$,
that we calculated in the following way.
We computed for a grid of points in the three-dimensional
$\tan^2\vartheta_{12}$--$\Delta{m}^2_{21}$--$c^2_{23} c^2_{24}$
parameter space  
the expected values of the observables and with those and the corresponding 
uncertainties we constructed the function  
$\chi^2(\tan^2\vartheta_{12},\Delta m_{12}^2,c_{23}^2c_{24}^2)$.  
We found its minimum,
$\chi^2_{\mathrm{min}}$,
in the full three-dimensional space considering as  
a unique framework both MSW and vacuum oscillations. The allowed
three-dimensional regions  
for a given CL are then defined as the set of points satisfying  
the condition 
\begin{equation} 
\chi^2(\tan^2\vartheta_{12},\Delta m_{12}^2,c_{23}^2c_{24}^2) 
-
\chi^2_{\mathrm{min}}
\leq
\Delta\chi^2(\mathrm{CL},3\,\mathrm{dof})
\label{deltachi2} 
\end{equation}  
where
$\Delta\chi^2(\mathrm{CL},3\,\mathrm{dof}) = 6.25, 11.36$
for 
CL=90\%, 99\%, respectively.
In  
Fig.~\ref{four} we plotted the  
sections of such volumes
in the
$\tan^2\vartheta_{12}$--$\Delta{m}^2_{21}$
plane
for six selected values of
$c^2_{23} c^2_{24}$.

\begin{figure}[t]
\begin{center}
\includegraphics[bb=8 10 238 220,width=\linewidth]{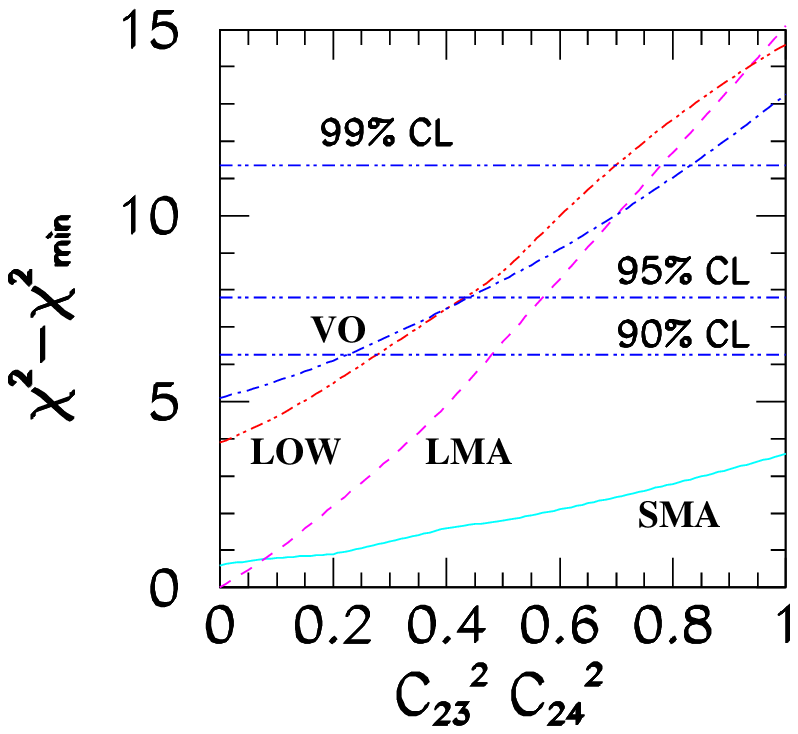}
\refstepcounter{figure}
\label{chi}
\\
\small
Figure \ref{chi}
\end{center}
\end{figure}

The 
global minimum of $\chi^2$,
$\chi^2_{\mathrm{min}} = 28.8$
with
23 degrees of freedom
corresponding to 28 data points
(4 rates
plus 6 Super-Kamiokande zenith-angle bins
plus 18 Super-Kamiokande energy spectrum bins)
minus 2 normalization factors
(for the Super-Kamiokande zenith-angle and energy spectrum data)
minus 3 fitted parameters,
lies in the LMA region for
$\cos^2{\vartheta_{23}} \cos^2{\vartheta_{24}} = 0$,
corresponding to pure
two-generation $\nu_e\to\nu_a$ oscillations,
and
$\tan^2\vartheta_{12} = 0.35$,
$\Delta{m}^2_{21} = 2.3 \times 10^{-5} \, \mathrm{eV}^2$.
However,
as one can see from Fig~\ref{chi},
the $\chi^2$ of the best-fit point
in the LMA, LOW and VO regions
rises steeply when
$\cos^2{\vartheta_{23}} \cos^2{\vartheta_{24}}$
is increased.
Therefore,
the 99\% CL LMA, LOW and VO regions disappear for
$c^2_{23} c^2_{24} \simeq 0.78, 0.70, 0.83$,
respectively.
For values of
$c^2_{23} c^2_{24}$
close to one,
corresponding to almost pure
two-generations $\nu_e\to\nu_s$ transitions,
only the SMA region is allowed.
This is expected, because
only the SMA solution is allowed
in the two-generation $\nu_e\to\nu_s$ fit of solar data
(see \cite{Gonzalez-Garcia-atm-analysis-99}).
Notice, however, that the statistical analysis is different: 
in the two-generation picture the pure
$\nu_e$$\to$active and $\nu_e$$\to$sterile 
cases are analyzed separately, 
whereas in the four-neutrino picture they are taken into account 
simultaneously in a consistent scheme that allows to calculate 
the allowed regions 
with the prescription given in Eq.~(\ref{deltachi2}). 
We think that the agreement between the results of the analyses 
with two and four neutrinos indicate that the 
physical conclusions are quite robust. 

The disappearance of the LMA and LOW regions
for values of
$c^2_{23} c^2_{24}$
close to one
is due to the following reason.
Unlike active neutrinos which lead to events in the 
water Cherenkov detectors by interacting via neutral current with the 
electrons, sterile neutrinos do not contribute to the Kamiokande and  
Super--Kamiokande event rates.  Therefore a larger survival probability  
for $^8\mathrm{B}$ neutrinos is needed to accommodate the measured rates. As a  
consequence a larger contribution from $^8\mathrm{B}$ neutrinos to the Chlorine  
and Gallium experiments is expected, so that the small measured rate in  
Chlorine can only be accommodated if less $^7\mathrm{Be}$ neutrinos are present in the 
flux. This is only possible in the SMA solution region, since in the 
LMA and LOW regions the suppression of $^7\mathrm{Be}$ neutrinos is not enough
(the survival probability of $\nu_e$'s
as a function of neutrino energy is rather flat). 

Figure~\ref{four} shows that the LMA 
region at 99\% CL extends to high values of 
$\Delta{m}^2_{21}$, 
even above $10^{-3} \, \text{eV}^2$ 
for $c_{23}^2c_{24}^2 \lesssim 0.1$. 
Since the atmospheric mass squared difference 
$\Delta{m}^2_{\mathrm{atm}}$ 
lies between $10^{-3}$ and $10^{-2} \, \text{eV}^2$ 
(see \cite{SK-sun-lp99}), 
one may wonder if the solar and atmospheric mass squared differences 
may coincide 
and three massive neutrinos may be enough for the explanation of 
solar, atmospheric and LSND data. 
The answer to this question is negative, 
because 
in the high--$\Delta{m}^2_{21}$ 
part of the 99\% CL LMA region the mixing angle $\theta_{21}$ is large, 
$ 0.3 \lesssim \sin^2(\theta_{21}) \lesssim 0.7 $, 
and in this case disappearance of $\bar\nu_e$'s 
should be observed in long-baseline reactor experiments, 
contrary to results of the CHOOZ \cite{chooz} experiment. 
In other words, 
the results of the CHOOZ experiment, 
that have not been taken into account in the present analysis, 
forbid the part of the 99\% CL LMA 
region that extends above 
$\Delta{m}^2_{21} \simeq 10^{-3} \, \text{eV}^2$.
For this reason we cut the plots at this value. 

\section{Conclusions}
\label{conclusions}

We have considered the two four-neutrino schemes
A and B in Fig.~\ref{4schemesAB}
that are compatible with the results of all neutrino oscillation experiments.
These two schemes are completely equivalent
for neutrino oscillation experiments,
but have different phenomenology in
$\beta$ and neutrinoless $\beta\beta$ decay experiments
(see \cite{BGG-review-98,Giunti-Neutrinoless-99,BGGKP-bb-99}).

In general the four-neutrino schemes A and B
allow simultaneous transitions
of solar $\nu_e$'s into active $\nu_\mu$'s, $\nu_\tau$'s
and sterile $\nu_s$,
as well as
simultaneous transitions
of atmospheric $\nu_\mu$'s into active $\nu_\tau$'s
and sterile $\nu_s$'s.

We have fitted the data of solar neutrino experiments
in terms of neutrino oscillations,
that in the schemes under consideration
depend on three parameters,
$\Delta{m}^2_{21}$,
$\tan^2\vartheta_{12}$
and
$\cos^2{\vartheta_{23}} \cos^2{\vartheta_{24}}$.
The allowed regions of these three parameters are depicted in
Fig.~\ref{four},
from which one can see that the SMA region
is valid for all values of
$\cos^2{\vartheta_{23}} \cos^2{\vartheta_{24}}$
(\textit{i.e.}
any combination of
$\nu_e$$\to$active
and
$\nu_e$$\to$sterile transitions),
whereas the LMA, LOW and VO
solutions disappear for values of
$\cos^2{\vartheta_{23}} \cos^2{\vartheta_{24}}$
close to one,
where
$\nu_e\to\nu_s$
transitions are dominant.

Other authors have recently performed fits of the atmospheric neutrino data
in the framework of the two four-neutrino schemes A and B
\cite{Yasuda-nufact00,Lisi-nu2000}.
We expect that in the future a combined fit of solar and atmospheric
neutrino data will allow to constraint 
further the mixing of four neutrinos.

\acknowledgments
MCGC and CPG acknowledge partial support by the spanish DGICYT under 
grants PB98-0693 and PB97-1261, by the Generalitat Valenciana under grant
GV99-3-1-01 and by the TMR network grant ERBFMRXCT960090 of the 
European Union.


\end{document}